\documentclass[aps,prb,amssymb,showpacs,twocolumn,amsfonts]{revtex4}
\usepackage{tabularx,graphicx,float}
\usepackage{subfigure}
\usepackage{amsmath}
\usepackage{amssymb}

\usepackage{bm}
\usepackage{wasysym}
\usepackage{ulem}

\usepackage{color}
\usepackage{verbatim}

\newcommand{\bk}{\vec{\bf k}}
\newcommand{\bq}{\vec{\bf q}}
\newcommand{\ba}{\vec{\bf a}}

\newcommand{\br}{\vec{\bf r}}
\newcommand{\bR}{\vec{\bf R}}
\newcommand{\bt}{\vec{\pmb \tau}}

\newcommand{\bp}{\vec{\bf p}}

\begin{document}

\bibliographystyle{apsrev}
%\draft
%\twocolumn[\hsize\textwidth\columnwidth\hsize\csname@twocolumnfalse%
%\endcsname
\title{Effect of electron-electron interaction on the Fermi surface topology of doped graphene.}
\author{R. Rold\'{a}n, M.P. L\'{o}pez-Sancho and F. Guinea}
%\address{
\affiliation{{Instituto de Ciencia de Materiales de Madrid,
CSIC, Cantoblanco, E-28049 Madrid, Spain.\\}
}
\date{\today}
%\maketitle
\begin{abstract}

The electron-electron interactions effects on the shape of the Fermi surface 
of doped graphene are investigated. The actual discrete
nature of the lattice is fully taken into account. A
$\pi$-band tight-binding model, with nearest-neighbor hopping integrals,
is considered. We calculate the self-energy corrections at zero
temperature. Long and short range Coulomb interactions are included.
The exchange self-energy corrections for graphene preserve the trigonal warping 
of the Fermi surface topology, although rounding the triangular shape. 
The band velocity is renormalized to higher value.
Corrections induced by a local Coulomb interaction, 
calculated by second order perturbation theory, do
deform anisotropically the Fermi surface shape. 
Results are compared to experimental observations and to other theoretical
results.

\end{abstract}
%]
\maketitle

\section{Introduction}

 Since its discovery \cite{NG04, NG05, ZTSK05} graphene, a two-dimensional
(2D) single crystal thermodynamically stable formed by a
single-layer of carbon atoms ordered in a honeycomb lattice, has
been thoroughly investigated. It forms the basic block of carbon
nanotubes, fullerenes, graphite, and graphite intercalation
compounds. The 2D electronic properties are well described by a
$\pi$-band tight-binding (TB) model \cite{W47}. The valence and
conduction $\pi$-bands touch only at the six corners, {\bf K},  of
the 2D Brillouin zone (BZ); the degeneracy point
of the valence and conduction bands is often termed Dirac point. At
half-filling, undoped graphene, the Fermi level $E_F$ lies at the
Dirac point. The low-energy physics of a perfect graphene sheet is
described by the relativistic Dirac equation. The dispersion
relation turns up to be isotropic and linear near $E_F$. The
low-energy excitations of the system are Dirac fermions with zero
effective mass and a vanishing density of states at the {\bf K}
points. Because of these peculiarities, graphene is considered a
model system to investigate basic questions of quantum mechanics.
Due to its transport properties graphene is a promising material for
nanoelectronic applications\cite{CGPNG08}.

Improvements in experimental resolution have led to high precision
measurements of the Fermi surface (FS), and also to the extraction of the
many-body effects from the spectral function, as reported by angle
resolved photoemission spectroscopy (ARPES) experiments \cite{DHS03}
in different materials. The recent isolation of an atomic layer of
graphite, graphene, has renewed the interest in the physics of
the three-dimensional (3D) graphite 
and new aspects of the electronic properties of 3D graphite
have been observed with improved experimental techniques\cite{KE07}.

In a high resolution ARPES study of disordered graphite samples, coexistence of sharp
quasiparticle dispersions and disordered features was found
\cite{L05}, and was explained in terms of Van Hove singularities (VHS) in
the angular density of states. Later on, by using ARPES, the linear
and isotropic dispersion of the bands, near the three-dimensional BZ
corners ({\bf H} points) of graphite, has been directly observed
coexisting with parabolic dispersion bands \cite{L06}. The constant
energy maps taken near the {\bf H} point present circular shape from
$E_F$ to $-0.6$eV. This circular shape combined with the linear
dispersion found near the BZ corners {\bf H} suggests that the
dispersion shows a cone-like behavior near each  point {\bf H},
similar to that expected for graphene. The constant energy maps
start to deviate from the circular shape at $-0.9$eV, and at
$-1.2$eV a rounded triangular shape is observed. 
A linear energy dependence of the quasiparticle life-time has been as
well measured by ultrahigh resolution ARPES on high quality crystals
of graphite {\cite{SSTS07}}. The low-energy excitations seem to be
dominated by phonons, while those for higher energies are
characterized by the electron-hole pair creation\cite{SSTS07}. The
quasiparticle life-time had been studied as well by ultrafast time
resolved photoemission spectroscopy (TRPES) \cite{Xu96} and a linear
$\omega$ dependence was reported. Anisotropy of quasiparticle
life-times has been reported by TRPES in highly oriented pyrolytic
graphite (HOPG) as well as an anomaly in the energy dependence
between  $1.1$ and $1.5$eV, in the vicinity of a saddle point in
the graphite band structure \cite{Mo01}. 

A linear energy dependence
of the quasiparticle lifetime had been theoretically predicted for
graphite\cite{GGV96} neglecting the interlayer hopping.
Even including an interlayer hoping of $0.25$eV, an anomalous quasiparticle
life-time was obtained in graphite\cite{SCRBEL01} within the $GW$ approximation,
with a linear energy dependence along the $\bf {K \Gamma}$ direction
for energies well above the interlayer hoping.   
A discussion about the interlayer coupling strength in graphite can
be found in Refs.\cite{LVG07,KE07}, along side with an overview of recent 
experiments in both graphene and graphite samples.

In a combined ARPES and theoretical ab-initio quasiparticle study of
the $\pi$-band structure and the Fermi surface in graphite single
crystals, it is found that electron-electron correlation plays an
important role in semi-metallic graphite and should be taken into
account for the interpretation of experimental results \cite{Ru07}.
The electronic correlations renormalize the electronic dispersion
increasing the Fermi velocity. The equi-energy contours of the
photoemission intensity show trigonal warping (higher by a factor of about 1.5
if compared to graphene) around  the {\bf KH} direction of
the graphite 3D BZ, in both
Local Density Approximation (LDA) and TB-$GW$ calculations, from
$-0.1$eV. Correlation effects are found to be stronger as the energy increases and
differences between LDA and $GW$ results are more noticeable at $-0.4$eV than
at $-0.1$eV \cite{Ru07}.

The recently available 2D graphene samples have been  intensively
investigated. From the experimental point
of view graphene presents advantages with respect to other 2D
systems. Graphene can be controlled externally and  exposed to
vacuum therefore can be directly probed by different techniques
\cite{NG04, NG05, ZTSK05}. Electron-electron interactions in
graphene are expected to play an important role due to its low
dimensionality and many-body effects have received great attention.
The quasiparticle dynamics in graphene samples has been addressed by
high-resolution ARPES \cite{R07}. It was found that the conical
bands are distorted due to many-body interactions, which renormalize
the band velocity and the Dirac crossing energy $E_D$. Electron-hole
pair generation effects are important near the Fermi energy and
electron-phonon coupling contribution to the self-energy is also
important in the Fermi level region: an electron-phonon coupling
constant of $\lambda \approx 0.3$ is deduced with the standard
formalism\cite {R07}. Around $E_D$, electron-plasmon coupling is
invoqued to explain the peak in the imaginary part of the
self-energy ${\rm Im}\Sigma$, found just below $E_D$, whose width
(and intensity) scales with $E_D$. Although the three scattering
mechanisms contribute to strongly renormalize the bands, in Ref.
\cite {R07} it is claimed that the quasiparticle picture is valid
over a spectacularly wide energy range in graphene. More
recently, the doping dependence of graphene electronic structure has
been investigated by ARPES in  graphene samples at different dopings
\cite{R07b}. Upon doping with electrons, the Fermi surface grows in
size and deviates from the circular shape showing the trigonal
warping, finally evolving into a concave triangular shape. An
electron-phonon coupling to in-plane optical vibrations is proposed
to explain the experimental results. The electron-phonon coupling
constant, extracted from the data, presents a strong dependence on
$\bk$, with maximal value along the {\bf KM} direction. The presence
of a Van Hove singularity in the {\bf KMK} direction,
confirmed upon doping, could be a possible explanation of the
enhancement of the electron-phonon coupling\cite{R07b}. 

The layered nature of
graphite/graphene as well as the presence of the VHS in the density
of states near the $E_F$, which would enhance many-body effects,
make contact with the physics of cuprates superconductors. The
similarities between graphene and the cuprates have been already
noticed \cite{BK05} and the important role of many-body effects in
the basic physics of graphene has been investigated earlier
\cite{GGV93}.

ARPES  investigation of graphene samples epitaxially grown on SiC substrate,
have reported the observation of an energy  gap  of $\approx 0.26eV$
at the {\bf K} point \cite{La07}.
It is proposed that the opening of the gap is induced by the interaction with
the substrate on which graphene is grown, that breaks the A and B sublattice
symmetry \cite{La07}.

From the theoretical point of view, graphene offers many possibilities and
the electron-electron interaction induced  effects have been widely analyzed.
The self-energy have been object of special interest, since it gives
relevant information from fundamental properties. Furthermore, theoretical
results can be compared  to   recent ARPES reported self-energies.
Many approximations have been followed to investigate
the self-energy in graphene. 
The inelastic quasiparticle
lifetimes have been obtained within the $G_0W$ approximation \cite{HHS06}
with  a full dynamically screened Coulomb interaction. The scattering rates 
calculated for different carrier
concentrations are in good agreement with ARPES data from Bostwick $et\,\, al.$ \cite{R07}
without including phonon effects, contrary to the experimental interpretation
of the data \cite{R07}. The nature of the undoped and doped graphene
has been as well discussed theoretically in terms of the behavior of the
imaginary part of the self-energy \cite{SHT07}: a Fermi liquid behavior is
found for doped graphene while the zero doping case exhibits a quasiparticle
lifetime linear and a zero renormalization factor indicating that, close
to the Dirac point, undoped graphene behaves as marginal Fermi liquid, in agreement
with previous theoretical work\cite{GGV96,GGV99}.

By evaluating exchange and random-phase-approximation (RPA) correlation energies,
an enhancement of the quasiparticle velocities near
the Dirac point is found in lightly doped graphene  
taking into account the eigenstate chirality \cite{BBPAMc07}.
The role of electron-electron interactions in the ARPES spectra of a n-doped graphene sheet
has been theoretically investigated \cite{McD07} by evaluating the self-energy
within the RPA and turned out to be important when interpreting experimental data.
Recently \cite{M07,SHHT07} the validity of the RPA
in the calculation of graphene self-energy has been discussed. The approximation
is shown to be valid and controlled for doped graphene where the Fermi level is
shifted up or down from the Dirac point. The RPA  fails for 
undoped graphene, where $E_F$ lies  at $E_D$.
In most of the self-energy studies graphene is described by 
the massless Dirac equation in the continuum limit.

In this work, we calculate the  corrections
induced by the electron-electron interaction
on the electronic band structure. We focus  on the corrections
to the Fermi surface shape of  doped graphene.
The Fermi surface is one of the key features needed
to understand the physical properties of a material and its shape
provides important information. Due to the 2D character of graphene,
electronic  interaction  should be  important.
We consider the $\pi$-band tight-binding model taking into account
the discrete nature of the  honeycomb lattice in order
to investigate  doped graphene and possible correlation effects in the trigonal warped
topology of the Fermi surface.
We calculate first the exchange self-energy considering
long- and short-range  Coulomb interactions.
Besides the renormalization of the band velocities, a deformation of
the trigonal warped Fermi surface is found at this level.
The second-order self-energy induced by an onsite Coulomb interaction is
as well calculated. The corrections to  the Fermi surface shape
are found to be anisotropic.

The paper is organized as follows.
In section II the model is presented and we explain the
self-energy calculation method. Section III presents the results
of the calculation and  Section IV contains a  discussion of the
results compared to experimental data and to other theoretical results
and the main conclusions of the work.

\section{The method}
\subsection{The model for the graphene layer}

\begin{figure}[ht]
  \centering
\subfigure[]{\label{Red}\includegraphics[width=0.27\textwidth]{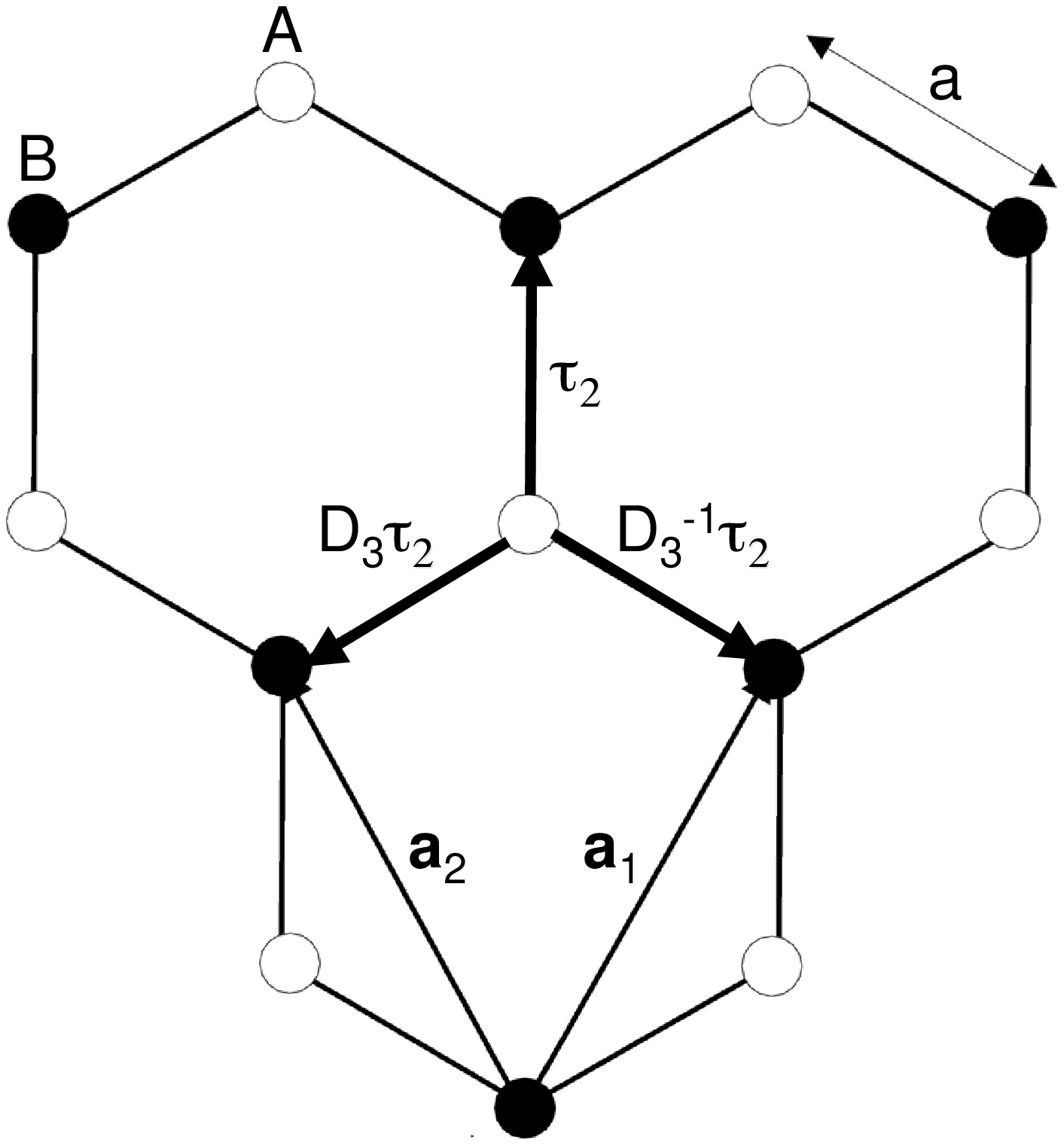}}
\subfigure[]{\label{BZ}\includegraphics[width=0.19\textwidth]{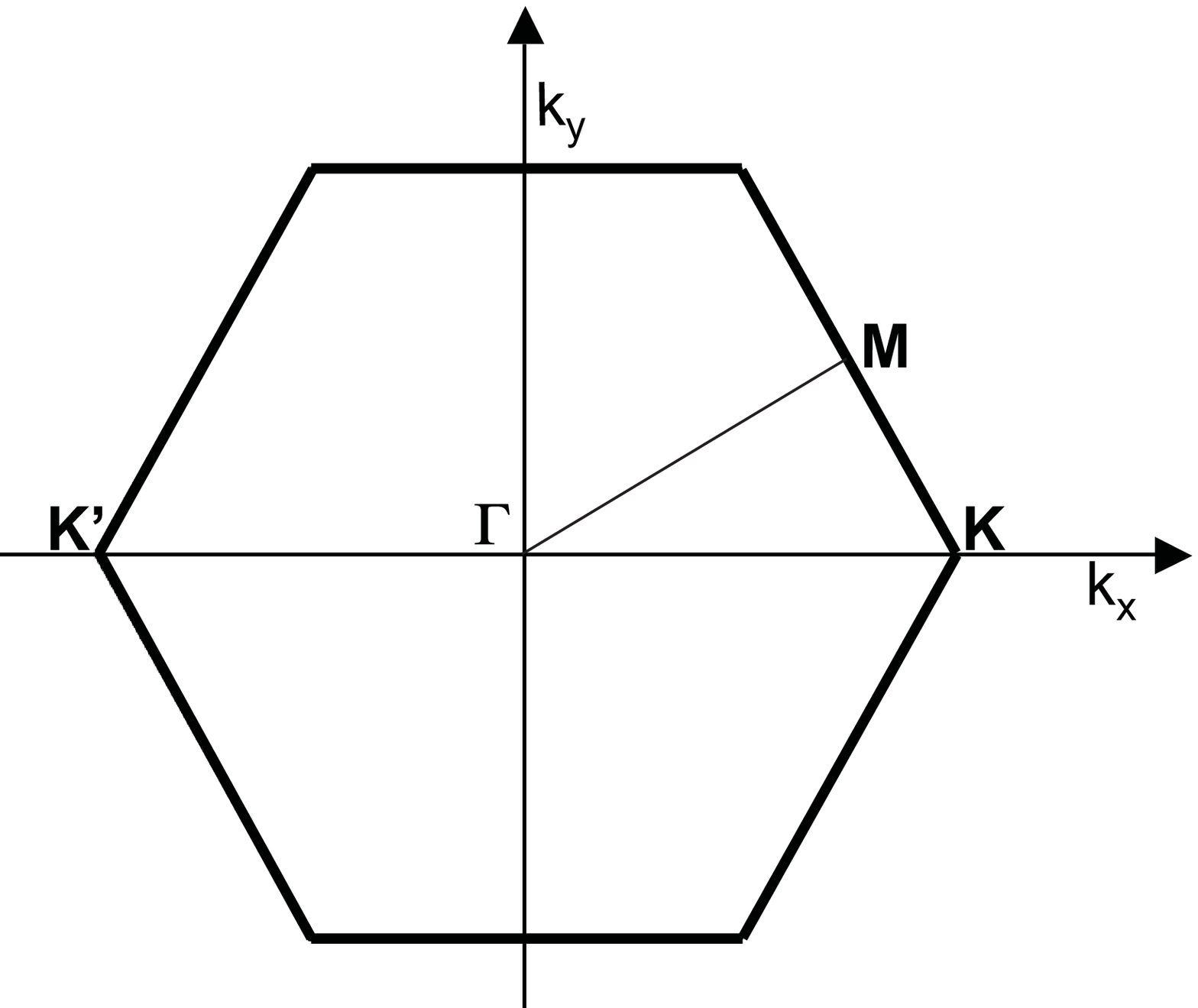}}
\caption{(a) Schematic representation of the structure of the honeycomb lattice,
open and solid  points represent the two inequivalent atoms.
The basic vectors $\ba_1$ and $\ba_2$ of the lattice are shown. The three  $\vec{\tau_i}$
connect   nearest neighbor atoms.  (b) First Brillouin zone of the honeycomb lattice.}
\label{RedBZ}
\end{figure}

Graphene is an atomic layer of carbon atoms arranged in a honeycomb
lattice with two atoms per unit cell, as shown in Fig.\ref{Red}. The distance
between nearest neighbor atoms is $a=1.42 {\rm \AA}$, and the primitive lattice vectors
are $\vec{\bf a}_1$ and $\vec{\bf a}_2$.
 The Brillouin zone is an hexagon, as
depicted in Fig.\ref{BZ}. We adopt the $\pi$-band tight-binding model
with only nearest-neighbor hopping\cite{W47}, since it captures the main
physics of the system as probed by more realistic models and by
experimental results.

The kinetic term of the Hamiltonian, considering only nearest-neighbor hopping, 
will be (we choose $\hbar=1$ throughout this paper)

\begin{equation}\label{H}
{\cal H}_{kin}=-t \sum_{\langle i  j \rangle,\sigma} {\hat a}^{\dagger}_{i,\sigma}{\hat b}_{j,\sigma} + h.c.
\end{equation}

where $t=2.82$eV is the nearest-neighbor hopping parameter, 
and the site energy of the $2p_z$ atomic orbital is taken as zero. 
The operator ${\hat a}^{\dagger}_{i,\sigma}$ (${\hat a}_{i,\sigma}$) 
is the creation (annihilation) operator
of an electron at site $ {\vec{\bf R}}_i$ on sublattice A with spin $\sigma$ ($\sigma=\uparrow,\downarrow$) 
(an equivalent definition corresponds to sublattice B).
By Fourier transformation to the momentum space we have

\begin{equation}\label{HK}
{\cal H}_{kin}(\bk)=t\left( \begin{array}{cc}
0 & g(\bk)\\
g^*(\bk) & 0\\
\end{array}\right)
\end{equation}

The function $g(\bk)$

\begin{equation}\label{g(k)}
g(\bk)=-\left( e^{-ik_ya}+2e^{ik_ya/2}\cos \left( \dfrac{\sqrt{3}}{2}k_xa\right) \right).
\end{equation}

is the structure factor of the honeycomb lattice.
Diagonalizing Eq.(\ref{HK})  the energy dispersion relation 
$\epsilon^0_{\lambda}(\bk)=\lambda t |g(\bk)|$ is obtained

%\begin{widetext}
\begin{equation}\label{dispersion}
\epsilon^0_{\lambda}\!(\bk)\!\!=\!\!\lambda t\sqrt{\!1\!\! + \!\! 4\cos\!\!\left(\! \dfrac{3}{2}ak_y\!\right)\!\! \cos\!\!\left(\! \dfrac{\sqrt{3}}{2}ak_x\!\right)\!\! +\!\!4\cos^2\!\!\left(\! \dfrac{\sqrt{3}}{2}ak_x\!\right)}
\end{equation}
%\end{widetext}

where $\lambda=-(+)$ for the valence (conduction) band. The two bands are degenerated
at the six corners of the BZ, $\bf{K}$ points. The corresponding Bloch wave functions are

\begin{equation}\label{wave-function}
\Psi_{\bk;\lambda}(\br)=\frac{1}{\sqrt{2}}\left( {\cal K}_{1\bk}(\br)+\lambda\frac{g^*(\bk)}{|g(\bk)|} {\cal K}_{2\bk}(\br)\right)
\end{equation}

\begin{figure}[t]
\centering
\includegraphics[width=0.4\textwidth]{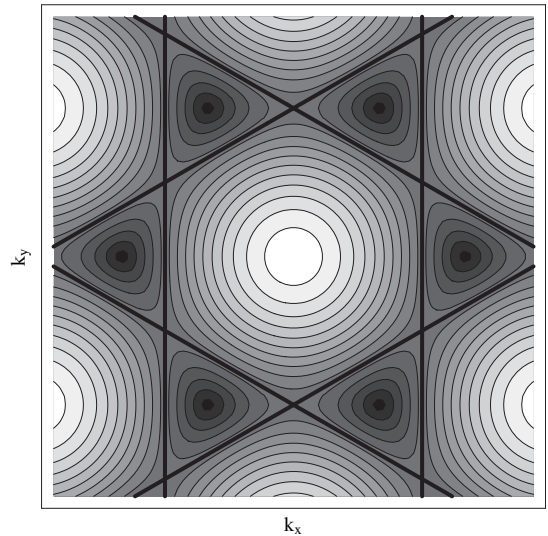}
\caption{Constant energy contours obtained with Eq.(\ref{dispersion}). The thick black lines corresponds to the Van Hove filling.}
\label{EC}
\end{figure}

The tight-binding functions are built from the atomic $2p_z$ orbitals\cite{B80} $\phi_z(\vec{\bf r})$

\begin{equation}
 {\cal K}_{i\bk}(\br)=\sqrt{\frac{A}{S}}\sum_{\vec{\pmb\rho}_n}e^{\imath\bk\cdot(\vec{\pmb{\rho}}_n+\vec{\pmb \tau}_i)}\phi_z(\br-\vec{\pmb\rho}_n-\vec{\pmb\tau}_i)
\end{equation}

where  $A$ and $S$ stand for the area of the unit cell and the
crystal respectively; $\vec{\pmb\rho}_n=n_1\vec{\bf a}_1+n_2\vec{\bf
a}_2$ is the lattice vector, and $\vec{\pmb\tau}_{1,2}$ define the
position of the two inequivalent atoms in the unit cell. We use
$\vec{\pmb\tau}_1=0$ and $\vec{\pmb\tau}_2=\frac{1}{3}(\vec{\bf
a}_1+\vec{\bf
  a}_2)$ (see Fig.\ref{Red}). At half-filling (undoped graphene) the Fermi
energy lies at the common point of the two bands (we take this
energy as our zero energy) and the Fermi surface is formed by
six points at the six BZ corners. The constant energy contours for the dispersion relation
Eq.(\ref{dispersion}) are  depicted in Fig.\ref{EC}. These six isolated points (only
two of them are inequivalent) are known as Dirac points because
around them, by a long-wavelength expansion, the kinetic energy term
of the Hamiltonian can be approximated by the 2D Dirac equation for
massless fermions. Upon doping, by following the constant
energy maps shown in Fig.\ref{EC},  the FS
points develop into circles and, eventually,   the FS adopt the 
rounded triangular shapes . As can be seeing in Fig.\ref{EC}, around 
each of the six BZ corners the energy lines are the same but rotated
with respect each other. Therefore, in the following, we will show 
the results in the vicinity of one of the corners of the BZ.

The electron interaction term of the
Hamiltonian ${\cal H}_{int}$ includes the Coulomb interaction
$e^2/\epsilon_0|\br_{ij}|$, where $|\br_{ij}|$ indicate 
 distances between sites of the honeycomb
lattice and $\epsilon_0 $ is  the dielectric
constant.  We will study the two limiting cases, long- and short-range Coulomb repulsion.
Among the short-range interactions, we consider 
interactions between electrons on neighbor atoms of the honeycomb lattice.

We also analyze corrections due to an on-site Coulomb repulsion $U$
between electrons with opposite spin on the same $p_z$ atomic orbital.
Although graphene is considered a weakly correlated system, the
effects of an on-site Coulomb interaction, Hubbard term,  on the
electronic properties of the honeycomb lattice have been
investigated in different scenarios. The phase diagram of the
Hubbard model in the honeycomb lattice has been studied by a
variety of techniques. Different instabilities have been found, from
the Mott-Hubbard transition \cite{ST92}, charge- and spin-density
wave \cite{TH92,LS01}, to superconductivity and magnetic phases
\cite{P02,H06,H07}. In our calculation we consider small values of
the on site repulsion, of the order of the hopping parameter for
graphene.

\subsection{Calculation of the self-energy}

\subsubsection{Long- and short-range Coulomb interaction: exchange self-energy.}

\begin{figure}[b]
\includegraphics[width=0.4\textwidth]{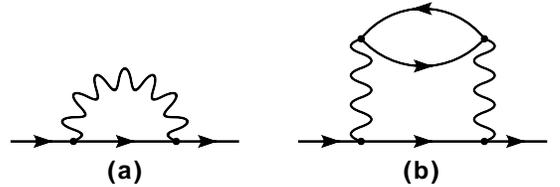}
\caption{(a) First-order self-energy diagram corresponding to the exchange contribution. (b) Second-order
self-energy  diagram.}
\label{diagrams}
\end{figure}

We calculate first the exchange self-energy contribution,
that corresponds to the one-loop diagram
shown in Fig.\ref{diagrams}(a).
The  self-energy has the form:

\begin{equation}\label{selfener}
\Sigma^x_{\lambda}(\bk,i\omega_n)\!\!=\!-\!\!\!\!\!\sum_{\lambda^{\prime}=\pm}\!\!\sum_{\bq}\!\!\frac{1}{\beta}\sum_{i\nu_n}G^0_{\lambda^{\prime}}(\bk+\bq,i\omega_n+i\nu_n)v_{\lambda\lambda^{\prime}}(\bk,\bk +\bq)
\end{equation}

where $\beta=1/k_BT$, $\omega_n$ and $\nu_n$ are fermionic and bosonic Matsubara frequencies respectively, and $G_{\lambda}^0(\bk,i\omega_n)$ is the bare
single-particle Green's function of an electron with momentum $\bk$
and band-index $\lambda$. The Coulomb
interaction matrix elements
$v_{\lambda\lambda^{\prime}}(\bk,\bk+\bq)$ between states
$|\bk,\lambda \rangle$ and $|\bk+\bq,\lambda^{\prime}\rangle$, are given by

\begin{widetext}
\begin{eqnarray}\label{vlamb}
v_{\lambda,\lambda^{\prime}}(\bk,\bk+\bq)&=&\int d\br_1\int d\br
\langle\Psi_{\bk,\lambda}(\br_1)|\langle
\Psi_{\bk+\bq,\lambda^{\prime}}(\br_1-\br)|\frac{e^2}{r}|\Psi_{\bk,\lambda}(\br_1-\br)\rangle|\Psi_{\bk+\bq,\lambda^{\prime}}(\br_1)\rangle\nonumber\\
&=&V(\bq)F_{\lambda\lambda^{\prime}}(\bk,\bk+\bq)
\end{eqnarray}
\end{widetext}

where $V(\bq)$ is the 2D Fourier transform of the Coulomb interaction  

\begin{equation}
V^{lr}(\bq)= \frac{2\pi e^2}{\epsilon_0 |\bq |}.
\end{equation}

We are also interested  in the effects of short-range Coulomb 
interactions between electrons on neighbor
atoms of the honeycomb lattice. The  Coulomb
interaction in momentum space between neighbor sites (matrix elements),  
described in Appendix \ref{ApendA}, can be expressed as

\begin{equation}\label{Vsr}
V^{sr}\!(\bq)\!=\! V\!\!\left (\!2\cos(q_ya)+4\cos\left(\frac{\sqrt{3}}{2}q_xa\right )\cos\left(\frac{1}{2}q_ya\right)\!\right).
\end{equation}

We take $V=t$ for the nearest-neighbor interaction strength.  
The function  $F_{\lambda\lambda^{\prime}}(\bk,\bk^{\prime})$ in Eq.(\ref{vlamb}) 
arises from the overlap of the wave functions obtained by diagonalizing 
the $\pi$-band tight-binding Hamiltonian\cite{B80,S86,LHC97},

\begin{equation}\label{solape}
F_{\lambda\lambda^{\prime}}(\bk,\bk+\bq)=\frac{1}{4} I^2(|\bq|)\left|1+\lambda\lambda^{\prime}\frac{g(\bk)g^*(\bk+\bq)}{|g(\bk)g(\bk+\bq)|}\right|^2
\end{equation}

Therefore the correct symmetry of the lattice is included in
$v_{\lambda\lambda^{\prime}}(\bk,\bk+\bq)$. $I(\bq)$ comes from the
matrix elements that contain the $2p_z$ wave function of carbon
atoms $\phi_z(r)$ and can be approximated by the unity\cite{S86}.
The function $F_{\lambda\lambda^{\prime}}(\bk,\bk+\bq)$,  close to the Dirac
points, reduces to the form,

\begin{equation}\label{fcont}
F_{\lambda\lambda^{\prime}}(\bk,\bk+\bq)= (1+ \lambda\lambda^{\prime}cos \theta_{\bk,\bk+\bq})/2
\end{equation}

where $\theta_{\bk,\bk+\bq}$ is  the angle between $\bk$ and
$\bk+\bq$. The form of the sublattice overlap matrix element for graphene, 
$F_{\lambda\lambda^{\prime}}(\bk,\bk+\bq)$, given in Eq.(\ref{fcont}) appears in the
theoretical studies based on graphene massless Dirac equation
continuum model \cite {HHS06, HS07, McD07,SHHT07,WSSG06}.

Considering the $T=0$ limit, after performing the summation of Matsubara frequencies, we can write

\begin{equation}\label{selfenerint}
\Sigma^x_{\lambda}(\bk)\!\!=\!-\!\!\!\!\sum_{\lambda^{\prime}=\pm}\!\!\int_{BZ}\frac{d\bq}{(2\pi)^2}V(\bq)F_{\lambda,\lambda^{\prime}}(\bk,\bk+\bq)\Theta(\mu-\epsilon^0_{\lambda^{\prime}}(\bk+\bq))
\end{equation}

where $\Theta$ is the Heaviside unit step function. The calculation
of $\Sigma^x_{\lambda}(\bk)$ requires a momentum integral over
the first BZ. Note that, since we are considering the discreteness
of the lattice, the $\bq$-integral does not have the  logarithmic
ultraviolet divergences appearing in the continuum model.
To carry out the integral in momentum space we divide 
the hexagonal BZ in two regions: the central region with small momenta
$\bq$, where the contribution of the long-range interaction is important,  and  the rest
of the BZ with larger momenta, where  short-range interactions,
namely, nearest-neighbor interactions,  are dominant. In the boundary
between both regions the potential functions are smoothly matched.
Details are given in Appendix B. As stated above, the matrix
elements $F_{\lambda\lambda^{\prime}}$ corresponding to intraband
($\lambda=\lambda^{\prime}=+$ and $\lambda=\lambda^{\prime}=-$) and
interband ($\lambda=+$ and $\lambda^{\prime}=-$) excitations are
calculated by Eq.(\ref{solape}).  Finally the correction of the
dispersion relation due to the interactions can be calculated from:

\begin{equation}\label{rendispersion}
\epsilon_{\lambda}(\bk)=\epsilon^0_{\lambda}(\bk)+\Sigma^x_{\lambda}(\bk)
\end{equation}

where $\epsilon^0_{\lambda}(\bk)$ was given in Eq.(\ref{dispersion}).  

>From Eq.(\ref{rendispersion}) we  obtain the corrections to the FS 
shape for  finite values of the doping.  
Notice that we address the corrections to the conduction band, 
taking into account the lattice symmetry.

\subsubsection{Local Coulomb interaction: second-order self-energy.}

In order to study the effects induced by local interactions in the
Fermi surface topology we compute the second-order self-energy. 
There are two diagrams that renormalize the
one-particle Green's function up to second-order in perturbation
theory. The Hartree diagram  gives a contribution
independent of momentum and energy and, hence, does not deform
the FS. The two-loop diagram, depicted in Fig.\ref{diagrams}(b),
modifies the FS topology through its $\bk$
dependence and changes the quasiparticle weight through its  $\omega$
dependence. 

We follow here the method explained in ref.\cite{R06c},
studying the interplay between the electron correlation and the FS
topology. We assume that the effect of high-energy excitations on
the quasiparticles near the FS is integrated out, leading to a
renormalization of the parameters of the Hamiltonian. Since we are
interested in low-temperature and low-energy processes, only the
particle-hole excitations within the energy scale about the Fermi line
defined by the cutoff $\Lambda$, are taken into account. We
compute the second order self-energy assuming that a FS, dressed by
the corrections due to the momentum independent interaction $U$, can be
defined. We consider electron doped graphene, thus the FS lies at
the conduction band. No singularities are reached for the values of the doping
considered.

We will perform an analytical calculation of the second-order self-energy.
We begin by momentum expanding the function  $g(\bk)$ given in Eq.(\ref{g(k)}), 
around two inequivalent Dirac points, obtaining

\begin{equation}
g(\bk)\approx \left\lbrace \begin{array}{ccc}
\frac{3}{2}a(k_x+\imath k_y)-\frac{3}{8}a^2(k_x-\imath k_y)^2 & {\rm around} & {\rm K}\\
-\frac{3}{2}a(k_x-\imath k_y)-\frac{3}{8}a^2(k_x+\imath k_y)^2 & {\rm around} & {\rm K}^{\prime}\\
\end{array}\right.
\end{equation}

where ${\bf \rm K}=\left(4\pi/3\sqrt{3}a,0\right)$ and 
${\bf \rm  K}^{\prime}=\left(-4\pi/3\sqrt{3}a,0\right)$. After diagonalizing 
the non-interacting Hamiltonian, we find the simplified dispersion relation

\begin{equation}
\xi^0_{\lambda,\nu}(\bk)\approx \lambda\frac{3}{2}at\sqrt{k^2+\frac{1}{16}a^2k^4-\nu\frac{1}{2}a(k_x^3-3k_xk_y^2)}
\end{equation}

where again $\lambda=\pm$ for the conduction and valence band,
$\nu=\pm$ for valley $\bf{K}$ and $\bf{K^{\prime}}$ respectively,
and $k=|\bk|$. To  second order in perturbation theory the renormalized
FS is given by the  solution of the equation:
\begin{equation}
\mu-\xi^0_{\lambda,\nu} (\bk)-{\rm Re} \Sigma^{(2)}_{\lambda,\nu} ( \bk , \omega=0 )=0
\label{dFS}
\end{equation}
where ${\rm Re} \Sigma^{(2)}_{\lambda,\nu} ( \bk , \omega )$ is the real part of
the self-energy of an electron of the  $\lambda$ band  with momentum $\bk$ and valley
index $\nu$. The  ${\rm Re} \Sigma^{(2)}_{\lambda,\nu} ( \bk , \omega )$ for 
a momentum independent interaction as the Hubbard $U$ 
has been computed from the imaginary part of the self-energy  following the method
explained in \cite{R06c} and turns up to be,

\begin{equation}\label{sigmaVH}
{\rm Re}\Sigma_{\lambda,\nu}(\bk,\omega=0)=-\frac{3}{32}\frac{U^2a^4}{\sqrt{2}\pi^3}\frac{\Lambda^2 {\rm sgn}(b(\bk))}{v_F^3(\bk)b(\bk)}
\end{equation}

where the frequency integral has been restricted to the interval $0
\le \omega \le \Lambda$, $\Lambda$ being the high energy cutoff taken
of the order of $t$. The Fermi velocity $v_F(\bk)$ and the curvature $b(\bk)$ 
of the non interacting FS are  given by the first and
second derivative of $\xi^0_{\lambda,\nu}(\bk)$ with respect to the
momentum, respectively. Both, $v_F(\bk)$ and $b(\bk)$,  are related to the parameters of
the Hamiltonian. For consistency we consider a weak local
interaction  $U \lesssim \Lambda$,  below the
energy cutoff. Forward and backward scattering channels are considered in the calculation. 
As explained in \cite{R06c}, this method of calculation of the
self-energy corrections does not depend on the microscopic model
used to obtain the electronic structure and the FS. Simple
analytical expressions of the effects induced by the interactions
are deduced from local features of the Fermi surface.

\section{Results}

\subsection{Corrections induced by the exchange self-energy}

\begin{figure}
  \centering
  \subfigure[]{\label{FSFull2-4}\includegraphics[width=0.22\textwidth]{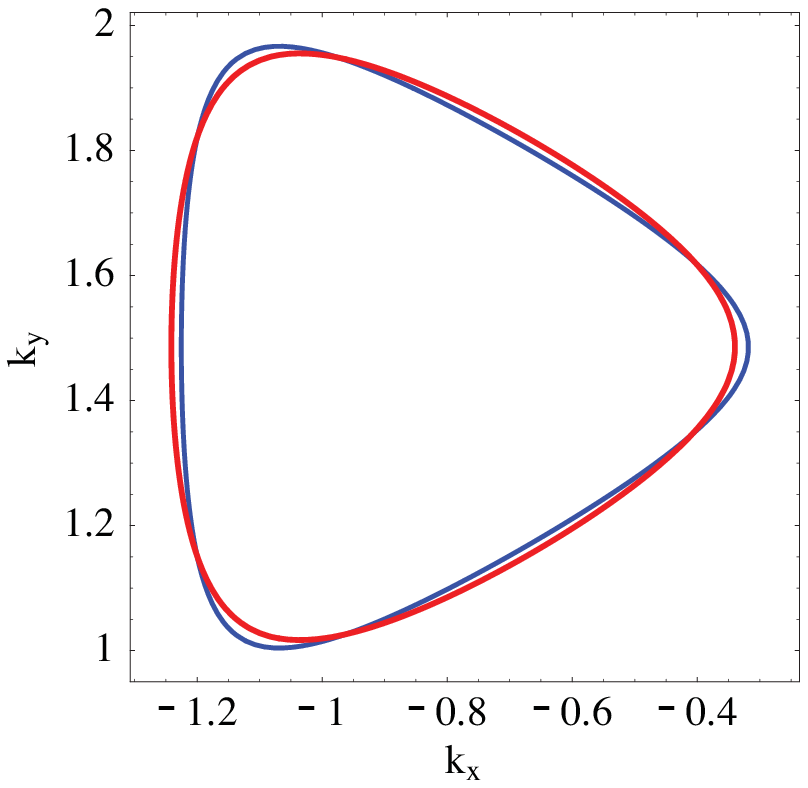}}
 \subfigure[]{\label{vFRenorFull}\includegraphics[width=0.22\textwidth]{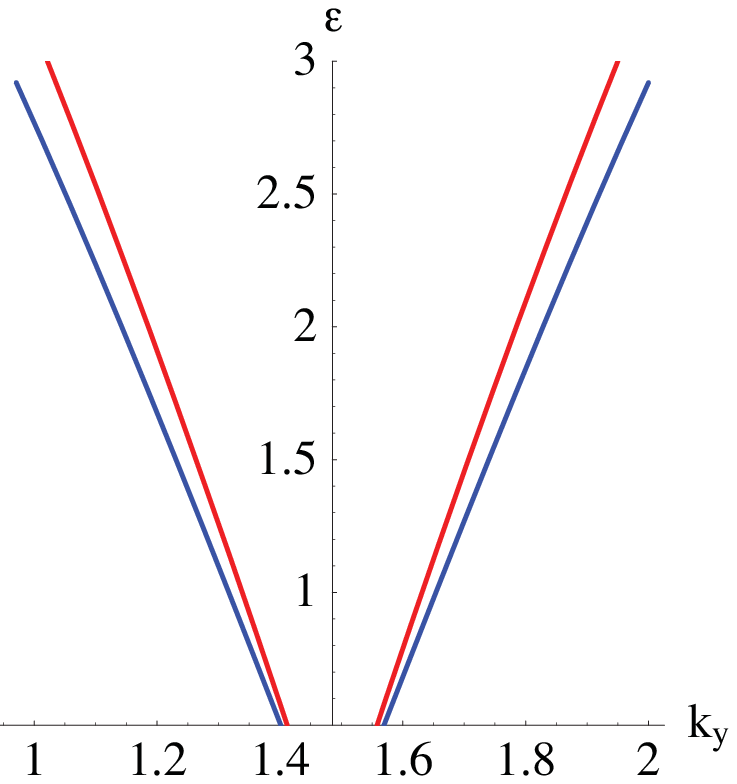}}
  \caption{Color online. (a) Non-interacting (blue) and interacting (red) Fermi lines around the $\bf{K}$ point for a doped graphene layer with $\mu\approx 2.4$eV due to the exchange self-energy. The origin of the coordinate system is at the $\Gamma$ point of the BZ. (b) Bare (blue) and renormalized (red) bands for $\mu=0.5$eV.}
  \label{SigmaFull}
\end{figure}

We now explain the numerical results obtained for the exchange self-energy calculated
from the formula Eq.(\ref{selfenerint}). As stated above, in the momentum integration
we consider the long-range Coulomb interaction in a  region
of the BZ around $\Gamma$,
and the nearest-neighbor interaction in the
rest of the BZ. The results show that the self-energy gives the stronger corrections
at the boundary lines of the BZ, keeping the  symmetry of the
lattice. The corrections to the FS topology are found to be small,
but not negligible,
as can be observed in Fig.\ref{FSFull2-4} for a doping density of $n\approx 5\times10^{14}$ electrons per cm$^{2}$
corresponding\cite{dopingd} to  a $\mu=2.4$eV. The self-energy corrections enhance the curvature of the 
sides  and round  the vertices of the triangular FS. This result agrees 
with the renormalization of the warping term found in Ref. \cite{AKT07} 
within one-loop renormalization group. One of the conclusions
reached in \cite{AKT07} is that the Coulomb interaction tends to suppress the warping term
making the energy surfaces more isotropic.

The self-energy effects are more noticeable in the band slope,  as can be observed in Fig.\ref{vFRenorFull},
where the dispersion relation, calculated  without and including the self-energy corrections, is shown.
The exchange self-energy enhances the velocity of the bands by a $20\%$ for the used parameter 
values ($V=t=2.82$eV), renormalizing the kinetic energy. 
The enhancement of the velocity has been obtained \cite{SHT07, GGV99, BBPAMc07, McD07}
previously. Calculating the self-energy within the on-shell approximation, a renormalization
of the the velocity compatible with Fermi liquid behavior was obtained for doped graphene \cite{SHT07}.
The increase of the velocity in lightly-doped graphene  has been attributed to the loss in
exchange energy when crossing the Dirac point, switching the quasiparticle chirality,
by evaluating exchange and RPA correlation energies\cite{BBPAMc07,McD07}.
All these theoretical works are based on the massless Dirac model for the low-energy excitations
of graphene.
We find that the exchange energy effects on the FS topology are  small for  doped 
graphene, in good  agreement with the ARPES results about the evolution of the graphene FS shape
with doping\cite{L06, R07b}.

\subsection{Corrections induced by a local interaction}

\begin{figure}
  \centering
 \subfigure[]{\label{FSLF}\includegraphics[width=0.22\textwidth]{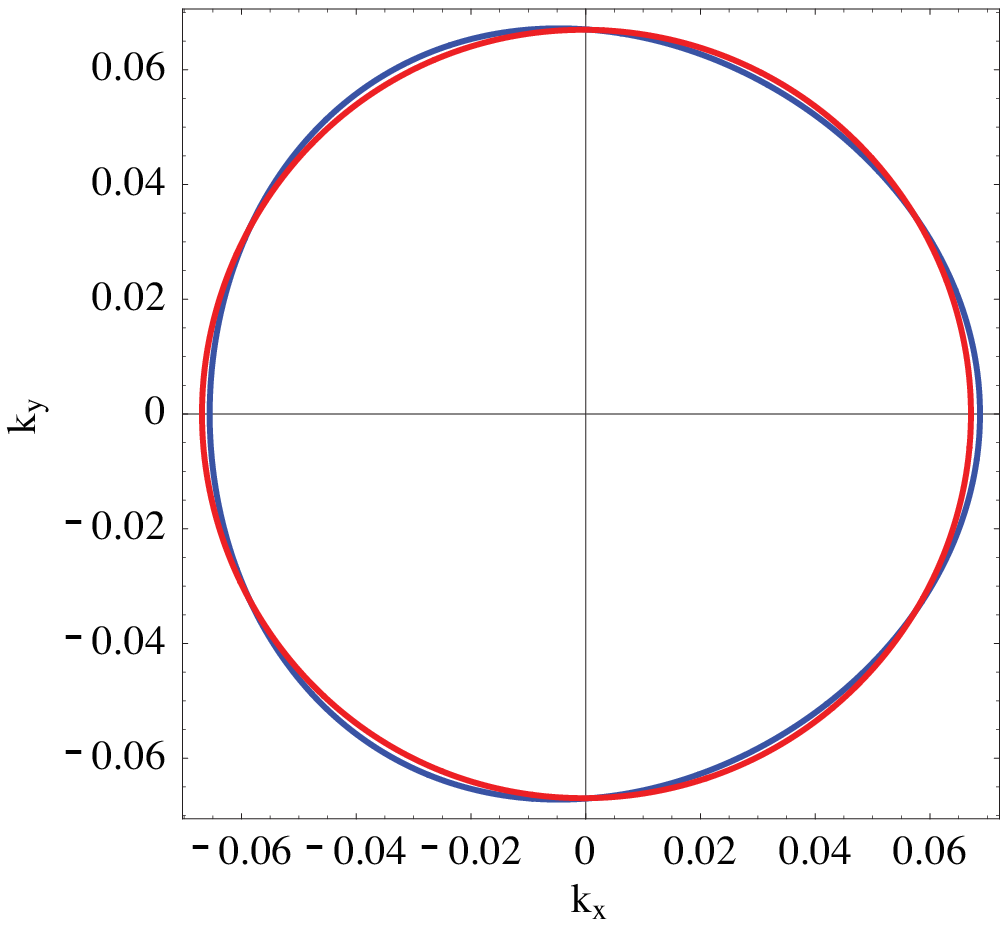}}
  \subfigure[]{\label{FSVH}\includegraphics[width=0.22\textwidth]{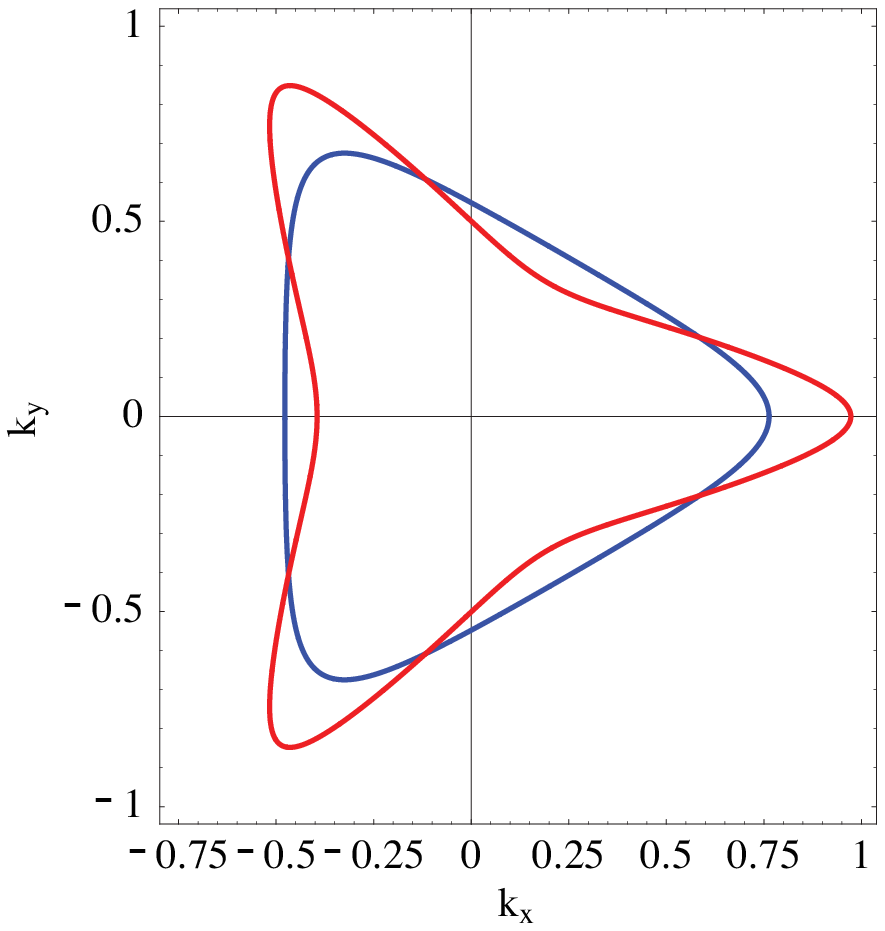}}
  \caption{Color online. (a) Non-interacting (blue) and interacting (red) Fermi lines around the $\bf{K}$ point 
for a doped graphene layer with $\mu\approx 0.2$eV due to a local Coulomb repulsion.  
(b) Same as (a) but for $\mu\approx 2.4$ eV. Notice the different scales of the coordinate axes. 
The origin of the coordinate system for this case is placed at the $\bf K$ point.}
  \label{SigmaLocal}
\end{figure}

In this section we analyze the corrections induced by a local interaction. 
The real part of the self-energy is computed from Eq.(\ref{sigmaVH}) which gives the second 
order perturbation theory renormalization of the Green's function.
As explained in Ref.\cite{R06c} this electron-electron interaction self-energy depends on
local features of the non-interacting FS, as the Fermi velocity and the curvature of the Fermi
line. We limit ourselves to the  weak coupling regime as seems generally accepted for graphene,
and where the perturbation approach is justified.
In Fig.\ref{SigmaLocal} the Fermi surfaces, at two different doping
levels, have been represented around the corner {\bf K} of the BZ. It should be notice
that the {\bf K'} counterpart has to be considered, (see Fig.\ref{EC}
where the  hexagonal
BZ is represented) in order to include all the possible scattering channels.

At low doping, for $\mu\approx 0.2$eV corresponding to a doping density of
about $3.9 \times 10^{12}$ electrons per cm$^{2}$, the FS has a circular shape  and the correction found
is small but appreciable.
The correction depends on $\bk$, as shown in Fig.\ref{FSLF}. By increasing the doping level
the FS adopts the round triangular shape. At a doping density of $n\approx 6\times10^{14}$ per cm$^{2}$, 
corresponding to a chemical potential of 
$\mu\approx 2.4$eV, the effects of the self-energy are noticeable,
and present a strong anisotropy. The deformation of the Fermi surface is maximum along the $\bf {KM}$ direction
of the BZ, as can be observed in Fig.\ref{FSVH}.
The topology of the FS is changed by the correction to a concave shape. The underlying
hexagonal symmetry ($2\pi/3$ rotation around $\Gamma$ point) is
preserved. This correction, with a maximum along the $\bf {KM}$ direction, is consistent
with the Renormalization Group analysis in\cite{GGV96b}, which shows that the
Van Hove point defines a stable fixed point, so that interactions deform the
Fermi surface towards the saddle point in the band dispersion.
Surprising enough, this deformation, induced by pure electron-electron
interaction self-energy,
presents a strong similarity with the deformation found by ARPES \cite{R07b}
in graphene. On Fig.3(a) from Ref.\cite{R07b} the plot of the Fermi contours derived from
curvefitting the
data, obtained at various  dopings, are presented.
The anisotropic deformation of the Fermi contours are  attributed
to electron-phonon coupling to the graphene in-plane optical vibrations.
The electron-phonon coupling constant $ \lambda_{\pi}(\bk)$ extracted from
experimental data
shows a strong anisotropy (around 5:1 ratio at highest doping) and a much higher strength
than would be expected for $\pi$-bands and optical phonons in graphene. It is argued
that the abruptness of the kink (change of the band velocity) and the broadening
of the bands suggest a strong mass renormalization by electron-phonon coupling.
The divergence of $ \lambda_{\pi}(\bk)$ along the $\bf{KM}$ direction is explained
as due to the VHS along the $\bf {KMK}$ boundary line of the BZ, which is reached 
upon further doping \cite{R07b}. The discrepancy between the experimental and theoretical  
electron-phonon coupling constants has been
addressed in ref. \cite {CM07} by using electron-phonon matrix elements extracted from density
functional theory simulations. It was found that, including finite resolution effects, the discrepancy between theory
and experiment was reduced from a factor of 5.5 to 2.2.

The presence of the strong
VHS reveals some similarity to the
cuprates\cite{D93}. The similarities between graphene and the cuprates have
been already remarked by Bena and Kivelson\cite{BK05} not only because of the
VHS but as well because of the similarity between the
band structure of graphite and that of the nodal quasiparticles in the
cuprate superconductors.
The existence of an extended VHS, as probed by the flatness of the bands in
both graphene and cuprates,
is important because of its effects on the divergences of the density of
states. The VHS is related
with the interactions which can be enhanced in its proximity and could increase
the superconducting critical temperature \cite{R07b}, 
in both  cuprates and in  graphite intercalated compounds,  
when $E_F$ is placed at the VHS.

In our calculated self-energy,
considering only a weak local electron interaction, the proximity to the VHS enhances its correction
and it reaches its maximum at the Van Hove band filling, where the Fermi level reaches the VHS 
($\mu= t$).
Therefore, at low doping, when the FS lies very close
to the $\bf K$ point, the corrections to the FS shape are very small since the
circular line lies far from the VHS location. On the other hand, at higher doping
the Fermi line lies close to the VHS,
specially the vertices of the triangle, and therefore the deformation is
higher in these directions. In Ref.\cite{R07b} by studying the evolution
of the coupling parameter $ \lambda_{\pi}(\bk)$ with the doping level
it is found that, while the minimum coupling strength grows only
slowly with doping, the maximum coupling parameter diverges as
the corresponding segment of the Fermi contour approaches the VHS at the
M point.

The corrections induced by the self-energy calculated
from Eq.(\ref{sigmaVH})  depend on local features of the non-interacting FS such as the Fermi
velocity $v_F$ and the curvature of the Fermi line. The main self-energy
effects in the square lattice were found to occur when the FS reaches the border of the BZ,
where the real part of the self-energy presents pronounced dips close
to the saddle points because the $v_F$ vanishes at the VHS\cite{R06c}. The same
behavior is found here for the honeycomb lattice. 

Because of the relation
between the real and imaginary part of the self-energy, a large self-energy correction
to the FS implies a large quasiparticle decay rate in the vicinity that region.

\section{Discussion and Concluding Remarks}
We have calculated the corrections induced by electron-electron
interactions in the band structure of graphene. We limit ourselves to the weak
coupling regime, as it is generally assumed to be appropriate for graphene.
At first order and zero temperature, the exchange self-energy induces a
deformation which contrarest the trigonal warping of the Fermi surface of
moderately doped graphene, in agreement with one-loop renormalization group 
calculations \cite{AKT07}.
The self-energy corrections round the
triangular shape. We find that the effect of the long-range Coulomb repulsion is
stronger at low dopings, when the Fermi level lies near the Dirac point. Due
to the circular shape of the FS at these fillings, the corrections to the
Fermi surface topology are small and only a renormalization of the chemical
potential is relevant at this level. On the other hand, the Fermi velocity is
enhanced.

The self-energy calculated by second-order perturbation theory, considering
a local Hubbard interaction, shows a strong $\bk$-dependence, therefore the
deformation induced in the FS topology is anisotropic. The  deformation is
highest
in the $\bf {KM}$ direction of the BZ, and the maximum value is reached near the Van Hove
filling, when the Fermi level is very close to the VHS ($\bf M$ point in Fig.\ref{BZ}).
This deformation of the FS shape  appears to be consistent with  the 
anisotropic deformation found
by ARPES attributed to a  coupling to the graphene in-plane
optical phonons\cite{R07b}. A different origin is proposed in \cite {CM07b}, where the
large and anisotropic values of the apparent electron-phonon coupling measured by ARPES
in graphene samples  doped with K and Ca \cite{R07b}, are explained by the strong
non-linearity in one of the $\pi^*$ bands below E$_F$ due to its hybridization with the
dopant Ca atoms\cite {CM07b}.
In the present work, it is induced by a purely electronic
interaction. 
The difficulties in the determination of the origin of some features
of the ARPES spectra  have been  discussed on a theoretical basis  in ref. \cite {CM07}.

We now comment on the connection between our work and recent theoretical work
on correlation effects on graphene. Most of the calculations
of the self-energy \cite{SHT07, GGV99, BBPAMc07, McD07} have been carried out  for undoped
or lightly doped graphene, in the continuum limit, where the conical shape of the bands
holds. The effect of the graphene lattice has been included in a recent
calculation of the self-energy
within the renormalized-ring-diagram approximation\cite{YT07}. 
Both doped and undoped graphene
present an imaginary part of the self-energy that, near the chemical potential, varies as
quadratic in the energy concluding that electrons in graphene behave
like a moderately correlated Fermi liquid\cite{YT07}. 

The role of electron-electron interactions in ARPES spectra of graphene has
been investigated within the RPA and it is shown to be important on the
$vk_F$ scale \cite{McD07} suggesting that, varying the carrier density, 
the effects of electron-electron interaction  can be separated from electron-phonon interaction
effects. The anisotropy of graphene energy-constant maps in ARPES has been 
proposed  to characterize quasiparticle properties \cite{MTGMF07} from
the electronic chirality on monolayer graphene to the magnitude and sign
of the interlayer coupling on bilayer graphene. Information about 
substrate-induced asymmetry may also be extracted from ARPES constant energy
patterns\cite{MTGMF07}.

Due to the high potential of photoemission to characterize graphene samples,
the knowledge of the many-body effects is crucial to interpret some feature
in the experimental spectra. The topology of the Fermi surface is related
with fundamental properties of the material. To keep track of the
different factors involved in the experimental data is  needed  
when comparing theory and experiment. As far as we know, corrections to the FS
topology of 2D doped graphene, the main focus of the present work, have not
been specifically addressed so far. We conclude that our results highlight the importance
of electronic correlation in  graphene.

{\it Acknowledgments.} Funding from MCyT (Spain) through grants
FIS2005-05478-C02-01 and from the European Union Contract
No. 12881 (NEST) is acknowledged. We are grateful to E. Rotenberg for
illuminating discussions. RR appreciates useful conversations with
A. Cortijo.

\appendix
\section{Coulomb interaction between nearest neighbor atoms in the honeycomb lattice.}\label{ApendA}

The   Coulomb interaction between electrons in nearest-neighbor atoms  in graphene
has the form

\begin{equation}
V(\br_i-\br_j)=\left\{\begin{array}{cc}
V=\frac{e^2}{a}\,\,\,&\,\,\, {\rm if} \,\,\, |\br_i-\br_j|=a\\
0\,\,\,&\,\,\,\,\,\, {\rm otherwise}\\
\end{array}\right.
\end{equation}

where $\br_{i,j}$ stand for lattice vectors in the two distinct A and B  sites of the honeycomb lattice.
Each atom of the sublattice A have three nearest neighbor atoms in the sublattice B,
and viceversa, $\br_j=\br_i+\bt_j$, where $\bt_j$ are vectors connecting 
nearest-neighbors sites on the sublattice B
from the sublattice A (see Fig.\ref{Red}). 
The nearest-neighbor interaction term of the Hamiltonian is given by (we omit spin indices)

\begin{equation}
{\cal H}_{V}=\frac{1}{2}\sum_{\bR}\sum_{j=1}^3V{\hat b}^{\dagger}_{\bR+\vec{\pmb \tau}_j}{\hat a}^{\dagger}_{\bR}{\hat a}_{\bR}{\hat b}_{\bR+\vec{\pmb \tau}_j}
\end{equation}

where $\bR=n_1\ba_1+n_2\ba_2$ is the position of an A atom in the lattice, with $n_{1,2}$ integer numbers. 
The above operators in the momentum space are  expressed  as

\begin{eqnarray}\label{FourierOperators}
{\hat a}^{\dagger}_{\bR}&=&\frac{1}{\sqrt{N}}\sum_{\bk \in BZ}e^{-i\bk\cdot\bR}{\hat a}^{\dagger}_{\bk}\nonumber\\
{\hat b}^{\dagger}_{\bR+\vec{\pmb \tau}_j}&=&\frac{1}{\sqrt{N}}\sum_{\bk \in BZ}e^{-i\bk\cdot(\bR+\vec{\pmb \tau}_j)}{\hat b}^{\dagger}_{\bk}\nonumber\\
\end{eqnarray}

where $N$ is the number of unit cells. The nearest-neighbor Coulomb interaction is found  to be

\begin{eqnarray}
{\cal H}_{V}&=&\frac{V}{2}\sum_{j=1}^3\sum_{\substack \bk,\bp,\bq\in BZ}e^{i\bq\cdot\bt_j}{\hat b}^{\dagger}_{\bk}{\hat a}^{\dagger}_{\bp}{\hat a}_{\bp-\bq}{\hat b}_{\bk+\bq}\nonumber\\
&=&\frac{V}{2}\sum_{\substack \bq\in BZ}f(\bq){\hat \rho}^{A}_{-\bq}{\hat \rho}^B_{\bq}\nonumber\\
\end{eqnarray}

where $f(\bq)=\sum_{j=1}^3e^{i\bq\cdot\bt_j}$ and the sublattice density operators are defined 
as ${\hat \rho}_{\bq}^{A}=\sum_{\bk\in BZ}{\hat a}^{\dagger}_{\bk+\bq}{\hat a}_{\bk}$ 
(an equivalent definition for ${\hat \rho}_{\bq}^{B}$). The operators given in Eq.(\ref{FourierOperators}) 
are related to the creation operators in the subband basis by the rotation 
${\hat c}^{\dagger}_{\bk,\lambda}=1/\sqrt{2}\left (e^{-i\theta_{\bk}/2}{\hat a}^{\dagger}_{\bk}+\lambda e^{i\theta_{\bk}/2}{\hat b}^{\dagger}_{\bk}\right)$. In this basis, the interacting Hamiltonian can be written as

\begin{equation}
{\cal H}_{V}=\sum_{\lambda ,\lambda^{\prime}=\pm}\sum_{\substack \bq\in BZ}V^{sr}(\bq){\hat \rho}^{\lambda}_{-\bq}{\hat \rho}^{\lambda^{\prime}}_{\bq}
\end{equation}

with the  expression given in Eq.(\ref{Vsr}) for $V^{sr}(\bq)$. 
Note that, in addition, the overlap matrix elements of the electronic wave-functions must be 
taken into account in the computation of the exchange self-energy.

\section{Integral in momentum space.}

\begin{figure}[b]
 \centering
\subfigure[]{\label{IntegReg}\includegraphics[width=0.27\textwidth]{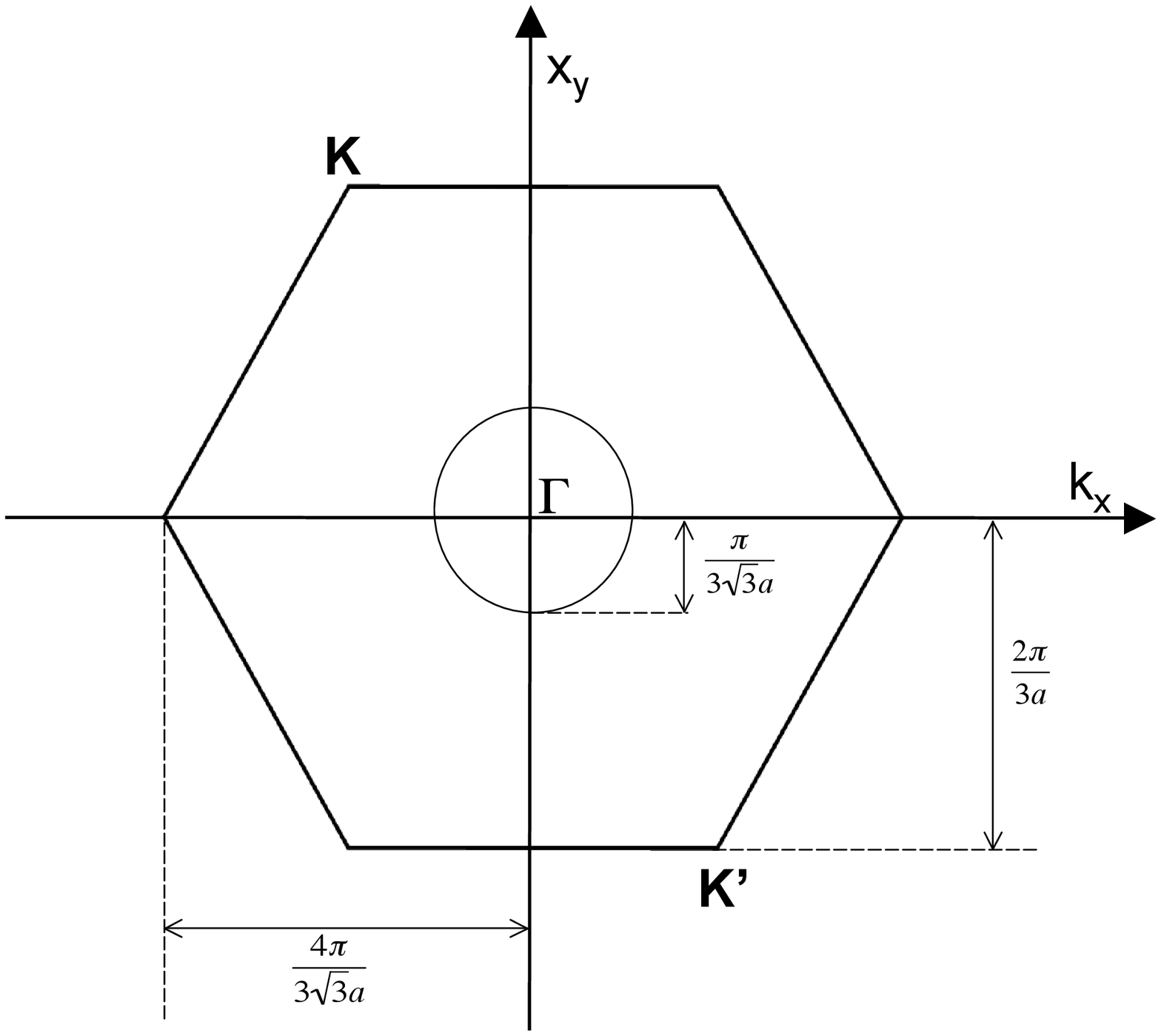}}
\subfigure[]{\label{CoulombFull}\includegraphics[width=0.20\textwidth]{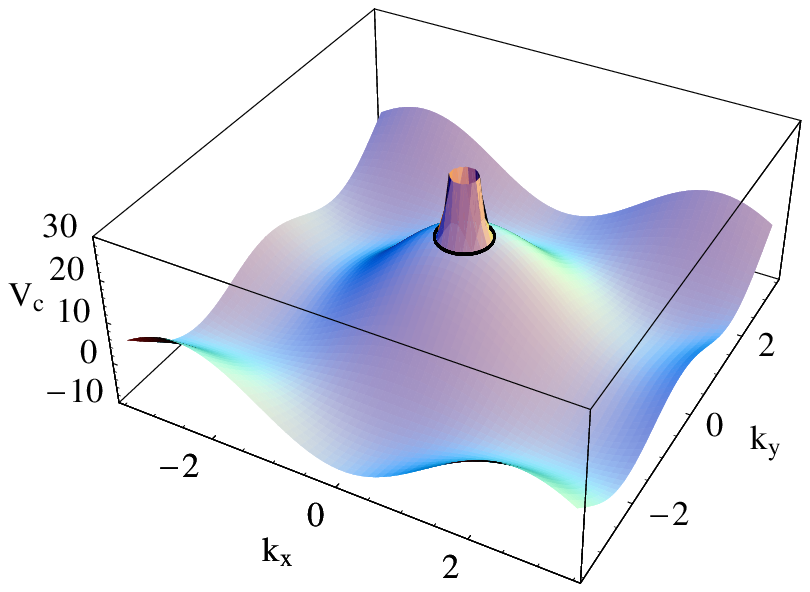}}
\caption{(a) Integration regions in the hexagonal Brilluoin zone. (b) Schematic plot of the Coulomb 
potential used in the calculation. The black circle separates the inner region 
where the long-range interaction is dominant  from the outer region where the short range-interaction dominates.}
\end{figure}

In order to calculate the exchange self-energy from Eq.(\ref{selfenerint}) we carry out a momentum
integral in the first Brillouin zone. We divide the BZ in two different regions as shown in Fig.\ref{IntegReg}.
The long-range interaction dominates in the central region while the nearest-neighbor interaction
is important in the outer region of the BZ. The circle centered in $\Gamma$, $0\leq |\bq|\leq \frac{1}{3}\frac{\pi}{\sqrt{3}a}$
is the boundary between the two regions.

 The interpolated potential considered in the calculation is schematically
plotted in Fig.\ref{CoulombFull}. The total potential, resulting from the 
combination of the long and short range interactions, is a continuous function consistent
with the lattice symmetry.

\bibliography{BibliogrGrafeno}

\begin{thebibliography}{48}
\bibitem[{dop()}]{dopingd}
\bibinfo{note}{The carrier density, for a given chemical potential, is
  calculated from $n(\mu)=1/A_c\int_0^{\mu}\rho(E)dE$ where $A_c=3\sqrt {3}
  a^2/2$ is the unit cell area and $\rho(E)$ is the density of states at the
  energy $E$. Our calculated values of doping densities and corresponding
  $\mu$, are in good agreement with the experimental data of the electron
  density reported for graphene chemically doped by potasium adsorption in
  Ref.\onlinecite{R07} where, for $\mu \approx 0.5eV$ the carrier density is of
  $n=1.1\times10^{13}$ per cm$^{2}$ and we found an electron density of
  $2.5\times10^{13}$ per cm$^{2}$ for the same value of $\mu$.}
\end{thebibliography}

 \end{document}